# Pillar Universities in Russia: The Rise of "the Second Wave"[1]

Tatiana Lisitskaya [*], Pavel Taranov[**], Ekaterina Ugnich[**] and Vladimir Pislyakov[***]

[*] *listsrostov@gmail.com*
Department of Finance and Credit, Faculty of IT and Economic Systems, Don State Technical University, Gagarin sq.1, Rostov-on-Don, 344000 (Russia)

[**] *ptaranov@donstu.ru; ugnich7746@gmail.com*
Department of World Economy and International Economic Relations, Faculty of Innovative Business and Management, Don State Technical University, Gagarin sq.1, Rostov-on-Don, 344000 (Russia)

[***] *pislyakov@hse.ru*
Library, National Research University Higher School of Economics, Myasnitskaya 20, Moscow, 101000 (Russia)

**Introduction. The system of Russian Universities**

The problem of identifying the leading universities in a country is rather easy to solve, one may focus, for example, on highly cited papers (e.g. Tijssen, Visser & van Leeuwen, 2002; Pislyakov & Shukshina, 2014; Abramo & D'Angelo, 2015) or other indicators of excellence. Sometimes it is more challenging to find the universities of "*the second wave*" which deserve to receive additional governmental help and budget because then they may become the most prominent ones and, so to say, enter the "Eredivisie", the highest football league. It is a more difficult task to find first among the seconds than to find the firsts among all.

Russian government tries to reform universities to more effectively invest public money into science-dependent technologies and innovations of the country (Schiermeier, 2010). "Economics of knowledge"—this is a goal of reforming and transforming of higher education in Russia (e. g. Dadasheva et al., 2016).

Today higher education institutions in Russia are classified as follows:
- Moscow State University and Saint Petersburg State University are the main National Russian Universities. In 2009 they were officially marked by status of the leading classical universities in Russia.
- From 2006 the "Federal Universities" were established. Now there are 10 such universities. They are deeply integrated with the regional governments and industries.
- There are 29 "National Research Universities", the program has started in 2009. Their mission is to combine education with scientific research. The idea is that this connection can help Russian higher education institutions to become the leading organizations on the market.
- The famous "5–100" project which was started in 2013 by the government (Rodionov, Yaluner & Kushneva, 2015; Ivanov, Markusova & Mindeli, 2016; Block & Khvatova, 2017; Guskov, Kosyakov & Selivanova, in print) now includes 21 universities. The aim is not only entering by, at minimum, five universities into the tops-100 of world universities' rankings (from there the name "5–100"), but also to evoke the "Triple Helix" (e. g. Leydesdorff, 2000) and to stimulate interaction between education, science and industry. Project "5–100" is not strictly linked to universities' classification and includes, among others, five Federal Universities and twelve National Research Universities.

---
[1] We thank Pavel Kasyanov (Clarivate Analytics) for his helpful consultations on Web of Science and InCites databases, Andrey Lovakov for his valuable comments on the draft of this paper.

At last, in 2015 RF Ministry of Education and Science has started to form a new type of higher education institutions, "*the Pillar Universities*" (sometimes also called in English "Flagship Universities", which is a reference to the US system of Flagship Universities (National Science Board, 2012, p. 26)). The aim of this project is to create strong educational and research centers specially oriented at the needs of regions. These institutions were not conceived as most prominent national leaders, but, indeed, as "pillars" of the whole higher education system which guarantee high-quality and reliable educational context all around Russia. They are not "leaders" but those who go right behind the leaders, *the second wave*.

The "pillar" status was granted in several stages. The main subject matter of the present paper are 11 pillar universities, those of the first stage which have received this status in the end of 2015. It is more reasonable to study only them because the first preliminary results of their progress already may be observed by bibliometric indicators. Moreover, from this analysis some cautious recommendations for the next stages of the project may be derived by experts in science policy. Now there are already 33 pillar universities in Russia and it is planned to increase their number to 100 by 2022.

Figure 1. Russian pillar universities of the first stage: geography.

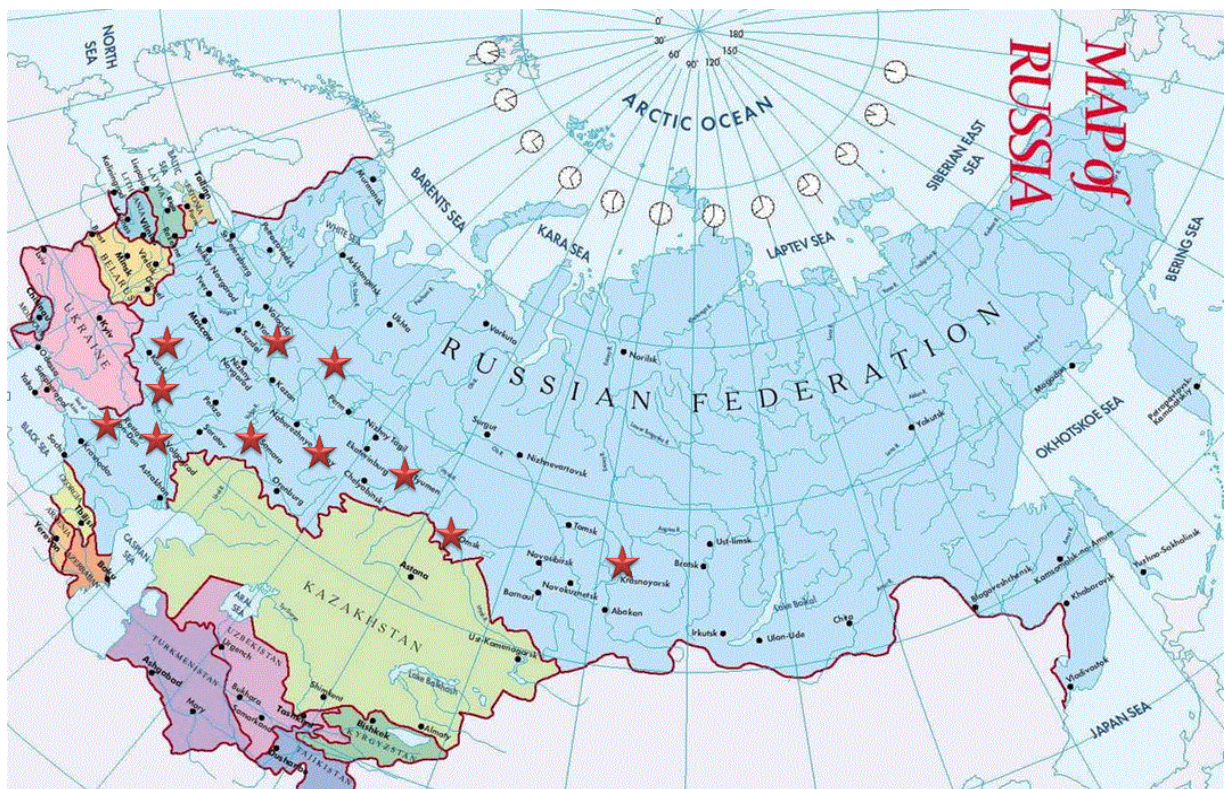

(source of the map template: avtocargo.by)

Pillar universities are regional centers of education, so further in the text we will denote the first stage participants by a city where they are situated:

- Volgograd = Volgograd State Technical University
- Samara = Samara State Technical University
- Tyumen = Tyumen Industrial University
- Voronezh = Voronezh State Technical University
- Vyatka = Vyatka State University
- Rostov = Don State Technical University (Rostov-on-Don)
- Omsk = Omsk State Technical University
- Krasnoyarsk = Siberian State Aerospace University (Krasnoyarsk)
- Ufa = Ufa State Petroleum Technological University
- Orel = Orel State University
- Kostroma = Kostroma State University

Geographical distribution of the pillar universities is visualized in Figure 1.

It should be noted that in the process of organizing of a pillar university 1–2 smaller or satellite or less independent universities of the region were incorporated into the main institution. Their budgets were also merged.

On average, the additional amount of public money each pillar university receives is from 100 to 150 mln rubles a year ($1.6–2.5 mln). As an estimate, it is about 4.5–6.7% of the average budget of these universities, though this value may greatly differ across institutions. The largest share of this extra budget was allocated to R&D (33%), accompanied with introduction of a new "effective contract system" with an evaluation framework for faculty and researchers. Additional funding is not very big, so the expert and information support from government as well as status of being 'pillar' itself play important roles.

Until now the start of the pillar universities project was analyzed either from reports of the institutions themselves, or from monitorings of the Ministry of Education and Science, or from local 'Interfax National university ranking' (Arzhanova et al., 2017; Surovitskaya, 2017). The aim of this paper is to study the pillars' performance using the internationally recognized bibliometric databases of Clarivate Analytics company (ex-Thomson Reuters IP).

**Data and Methods**
Web of Science (WoS) platform with its Science Citation Index Expanded (SCIE) and Social Sciences Citation Index (SSCI) databases were used as a source of bibliometric data. We omitted conference proceedings and book citation indexes because they contain different types of documents. Arts & Humanities Citation Index (AHCI) also was not included as humanities literature may not be assessed well by bibliometric indicators. Also, there are no impact factors (IF) for AHCI journals, while we will need IF data in our study.

We calculate indicators for two years, 2013 and 2017. It makes possible to track progress of the pillar universities in their process of becoming "pillar". Five years are enough to observe the evolution of the universities receiving their new status. Only "Article" and "Review" document types were considered, all other documents not taken into account.

As was explained in Introduction, the inauguration of the new type of universities in Russia, the "pillar universities", was accompanied by merging of several minor higher education institutions. That is, in 2017 one pillar university was a result of integration of universities being separate in 2013. To make bibliometric indicators comparable, we sum up all 2013 components of the future pillar university which is established later, in 2015.

**Results and Discussion**

*Output*
Figure 2 shows the progress of publication output of the organizations which have become pillar universities in the end of 2015. Remind that for the earlier period the data for all components of the future pillar universities are merged, so it is not an extensive growth simply caused by consolidating of the organizations. Generally, on the whole time interval the observed growth is close to linear. This means that not only after becoming "pillar" the universities have started their progress, but those institutions which were chosen for the first stage were already "ripe" and demonstrated success.

Figure 2. Papers by all 11 pillar universities (together), 2013–2017.

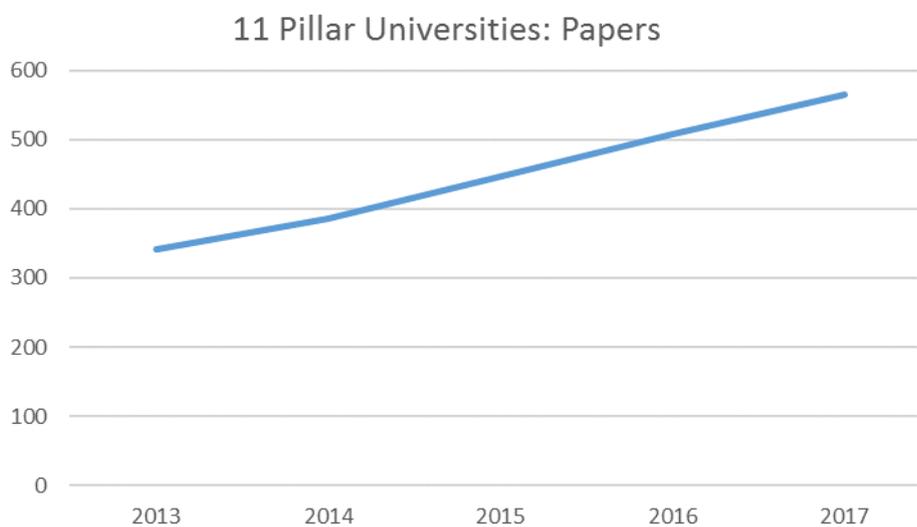

The total number of Russian WoS publications also grew during this period. But the progress of pillar universities even outperforms that of Russia. In 2013 the papers of these 11 universities had a 1.15% share in the total Russian output, while in 2017 this share has become 1.50%. Still, by now it is far less than, for example, the share of 21 universities from '5–100' project (33.6% in 2017).

The individual performance of each of the pillars is shown in Figure 3. Not one of the 11 has demonstrated a decline in publications in the prominent international journals. The leader has changed, instead of Volgograd in 2013, five years later the maximum number of WoS papers were published by pillar from Samara. What is probably even more important is that "weak" institutions of 2013, with no more than 10 publications in WoS, have the strongest progress: Kostroma (6→23) and Tyumen (10→39); both leave their last places taken in 2013 ranking.

Of course, to comprehensively compare the data we should make normalization by the number of faculty and staff in each of these universities. But these data are not easily available in Russian context and we leave this task for the development of our study. But in terms of relative growth, if we consider only universities with more than 10 papers in 2013 (to exclude outliers), the most striking progress shows Rostov (273%), Omsk (235%) and Vyatka (143%), which is equivalent to 39.0%, 35.3% and 24.8% of yearly growth respectively. The most moderate increase of output is found in Volgograd—only 4 extra paper (5.8%).

Figure 3. Papers by 11 pillar universities, 2013 and 2017.

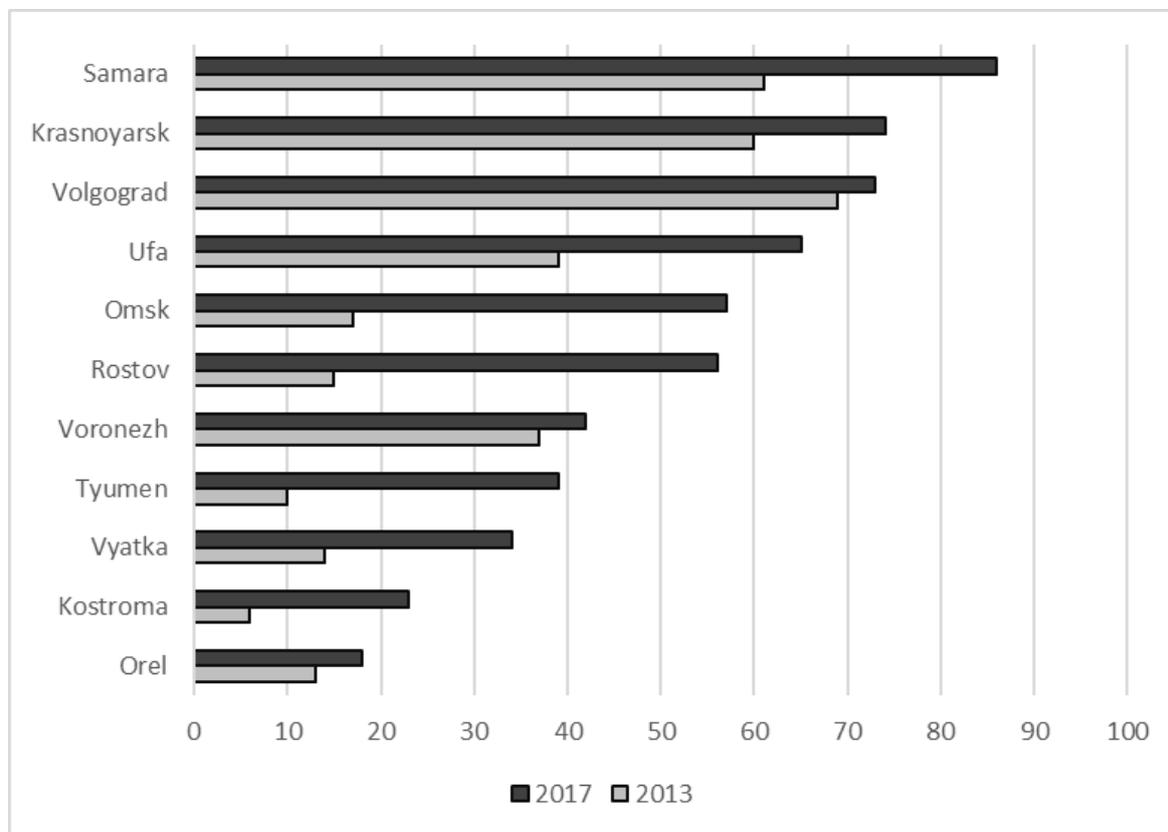

Some of the pillar universities were organized in the regions where already was a 'grand' university, either Federal University or a member of '5–100' program. Table 1 compares publication statistics and its dynamics for 'pillars' and 'grands'.

Table 1. Pillar universities and 'grand' (either Federal or 5–100) universities in the same region: Number of publications.

|  | 2013 | | 2017 | | growth, % | |
| --- | --- | --- | --- | --- | --- | --- |
|  | **pillar** | **grand** | **pillar** | **grand** | **pillar** | **grand** |
| Krasnoyarsk | 60 | 226 | 74 | 354 | 23 | 57 |
| Rostov | 15 | 299 | 56 | 429 | 273 | 43 |
| Samara | 61 | 96 | 86 | 248 | 41 | 158 |
| Tyumen | 10 | 43 | 39 | 173 | 290 | 302 |

'Grands': Krasnoyarsk—Siberian Federal University; Rostov—Southern Federal University; Samara—Samara National Research University; Tyumen—Tyumen State University.

Predictably, the 'grand' universities play a leading role, producing the largest number of papers in their regions. But in some cases (Rostov) the progress in output is more pronounced for a smaller pillar university.

*Impact*

In 2018 it is senseless to count citations received by papers published in 2017. Still, we may use as a proxy of a paper's quality the visibility of a journal where it was published. The latter may be estimated by IF from Journal Citation Reports database. Grančay et al. (2017) cite as critics of this approach (Seglen, 1997; Callaway, 2016), so the arguments in favor of it (Yuret, 2016). The authors of Grančay et al. paper conclude that "publishing in a journal with high IF is a certain mark of excellence." To assess multidisciplinary output of universities correctly, we divide journals into four quartiles in each WoS subject category. If a journal is assigned to different quartiles in different categories, the highest quartile is used.

Distribution of pillar universities' publications across journal quartiles is shown in Figure 4. To ensure consistency, the same year of Journal Citation Reports is used (2017). The progress made in five years is undoubted. The share of papers published in the lowest quartile has decreased. What is wonderful, is that the share of publications in the top-25% journals has grown by 2.4 times, from 5.3% to 12.5%! Only 7 out of 11 institutions succeeded to publish their papers in the highest quality journals in 2013. In 2017, each university has at least two papers in Q1. Vice versa, while in 2013 one university (Kostroma) published *all* its papers in Q4, in 2017 the maximum share of pillar's publications in the lowest quartile is 77% (Ufa).

Figure 4. Papers of 11 pillar universities, by Journal Citation Reports IF quartiles, %.

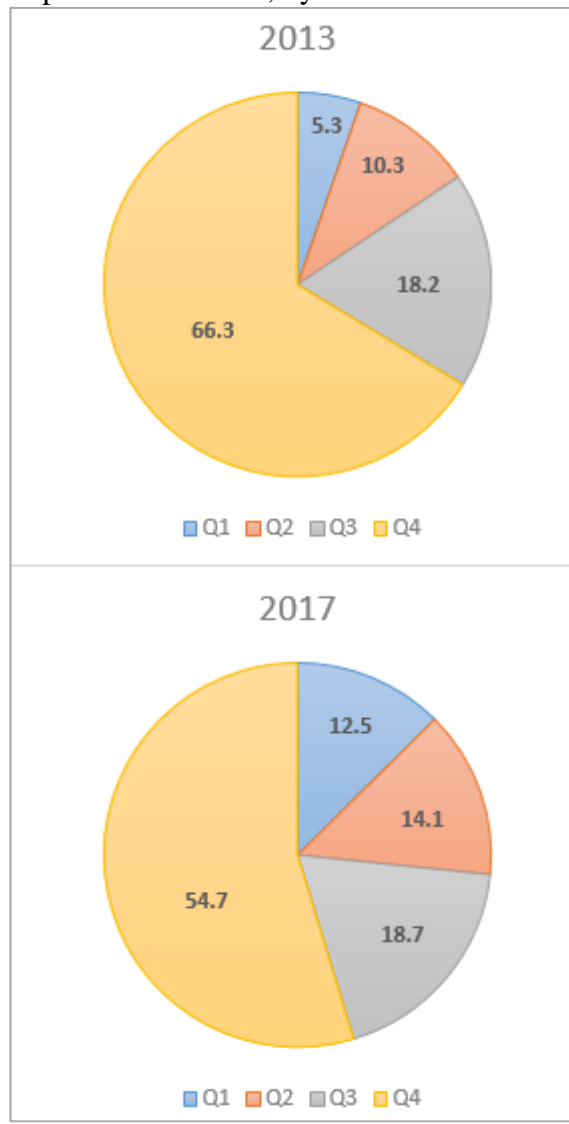

Though the advancements are striking, there is still a long way to go. According to InCites database, for the whole Russia the share of papers published in Q1 journals is 26% (year 2017). For Moscow State University, the biggest Russian higher education institution, this value equals 35%; for the biggest social sciences university, Higher School of Economics, 32%. Kostroma now has become a champion among pillars and it has 22% of its papers in the top journals (2017).

*Collaboration*

As is shown in Table 2, the intra-national ("domestic") collaboration of pillar universities has strengthened during five years 2013–2017. Almost all pillars started to write a greater percentage of papers in coauthorship with other Russian organizations. The choice of partners is most often based on geography, the most active partnership tends to occur between institutions of the same city. Still, there are some exceptions such as pairs Samara-Moscow, Vyatka-Kazan or Kostroma-Ivanovo (the latter are less than 100 km from each other). The leading partners of pillar universities are either other universities or institutes of the Russian Academy of Sciences, almost in an equal proportion.

Table 2. Domestic collaboration of pillar universities, share in the total output and most prolific organizations-coauthors.

| pillar university | 2013 (%) | 2017 (%) | main partner in 2017 (if more than 5 coauthored papers) |
|---|---|---|---|
| Samara | 18 | 43 | Peoples Friendship University of Russia (Moscow) |
| Krasnoyarsk | 83 | 89 | Kirensky Institute of Physics (Krasnoyarsk) |
| Volgograd | 43 | 37 | — |
| Ufa | 64 | 69 | Institute of Petrochemistry and Catalysis (Ufa) |
| Omsk | 59 | 82 | Institute of Hydrocarbons Processing (Omsk) |
| Rostov | 53 | 37 | Southern Federal University (Rostov) |
| Voronezh | 46 | 69 | Voronezh State University |
| Tyumen | 70 | 74 | Tyumen State University |
| Vyatka | 57 | 88 | Kazan Federal University |
| Kostroma | 17 | 61 | Institute of Solution Chemistry (Ivanovo) |
| Orel | 38 | 44 | — |
| **All together** | **48** | **62** | |

What is striking, it is that inter-pillar collaboration is almost zero. There were no papers coauthored by two pillar universities in 2013, and only two such papers in 2017 (both in the field of chemistry, Samara-Omsk and Samara-Ufa). This kind of collaboration is still waiting to be developed.

Table 3 contains share of papers created with foreign coauthors. Again, the majority of pillar universities have boosted their international collaboration (Volgograd and Voronezh are exceptions). While in 2013 pillars of Ufa, Vyatka and Kostroma had not a single paper in international partnership, five years later all 11 universities have an experience of research together with foreign colleagues. Interestingly, Rostov is the unique pillar which has more papers in international than in domestic collaboration in 2017. In five years it also reduced the share of its national co-authorship and rose the international one. One may say that Rostov pillar has refocused its collaboration efforts from domestic to international collaborations.

The percentage of internationally coauthored papers among pillar institutions taken as a whole has increased from 14% to 23%. It is still lower than all-Russian value of 39% for 2017. The leading partners of pillar universities among countries are the same as for the whole Russia—USA and Germany.

Though Surovitskaya (2017), based on the monitoring of the Ministry of Education and Science, states that international activity tends to lessen in 6 out of 11 pillar universities, our analysis clearly shows that bibliometric indicators of partnership with foreign colleagues grow for 9 institutions of the "first stage" pillar universities. The crucial role of such international collaboration for Russian institutions is shown by Pislyakov and Shukshina (2014) who study the most highly cited papers.

Table 3. International collaboration of pillar universities, share in the total output.

| pillar university | 2013 (%) | 2017 (%) |
|---|---|---|
| Samara | 13 | 16 |
| Krasnoyarsk | 18 | 35 |
| Volgograd | 12 | 8 |
| Ufa | – | 14 |
| Omsk | 12 | 25 |
| Rostov | 27 | 43 |
| Voronezh | 30 | 29 |
| Tyumen | 10 | 33 |
| Vyatka | – | 21 |
| Kostroma | – | 9 |
| Orel | 23 | 33 |
| **All together** | **14** | **23** |

**Conclusion**
Russian government has started to directly support not only "the best" higher education institutions, but also "the best among the seconds", as we call it in this paper "the second wave" universities. Too little time has passed to make comprehensive conclusions, but the first results are encouraging.

For the 11 pillar universities of the first stage of the project the publication output has increased more than 1.6 times compared to 2013, these papers being published in more visible journals (2.4 times growth in the number of publications in Q1 titles). These institutions become more and more involved into scientific networks as at domestic so at international level. International collaboration strengthens, now nearly each fourth paper by pillar university is written with foreign coauthor(s).

What is even more important, no 'total sleepyheads' remained. In 2017 each of the 11 first stage universities has published at least two papers in Q1 journal and at least two papers in international collaboration.

Some cautious policy recommendations also may be made. For example, it is needed to motivate collaboration between pillars themselves (and, consequently, between regions), an activity not developed so far. As a suggestion, some regular joint conferences (scientific, not purely administrative meetings) for all pillar universities could reinforce their cooperation.

Further development of the "second wave" governmental project may further evoke competition of pillars with the 'grands' of Russian university system. This will benefit both sides, as 'pillars', so 'grands'. And hopefully some of the most successful pillar universities could enter in the future the highest league of Russian education and those who are second will be first (cf. Mt. 19:30).

**References**

Abramo, G. & D'Angelo, C.A. (2015). Ranking research institutions by the number of highly-cited articles per scientist. *Journal of Informetrics*, 9(4), 915–923. DOI: 10.1016/j.joi.2015.09.001.

Arzhanova, I.V., Vorov, A.B., Derman, D.O., Dyachkova, E.A. & Klyagin, A.V. (2017). Results of pillar universities development program implementation for 2016. *University Management: Practice and Analisys*, 21(4), 11–21.

Block, M. & Khvatova, T. (2017). University transformation: Explaining policy-making and trends in higher education in Russia. *Journal of Management Development*, 36(6), 761–779. DOI: 10.1108/JMD-01-2016-0020.

Callaway, E. (2016). Publishing elite turns against impact factor. *Nature*, 535(7611), 210–211. DOI: 10.1038/nature.2016.20224.

Dadasheva, V., Efimov, V. & Lapteva, A. (2016). The Future of Higher School in Russia: Missions and Functions of Universities. In L.G. Chova, A.L. Martínez, I.C. Torres (Eds.), *Proceedings of INTED2016, Tenth International Technology, Education and Development Conference. 7–9 March 2016* (pp. 286–296). Valencia: IATED Academy. DOI: 10.21125/inted.2016.1072.

Grančay, M., Vveinhardt, J. & Šumilo, Ē. (2017). Publish or perish: How Central and Eastern European economists have dealt with the ever-increasing academic publishing requirements 2000–2015. *Scientometrics*, 111(3), 1813–1837. DOI: 10.1007/s11192-017-2332-z.

Guskov, A.E., Kosyakov, D.V. & Selivanova, I.V. (in print). Boosting research productivity in top Russian universities: the circumstances of breakthrough. *Scientometrics*. DOI: 10.1007/s11192-018-2890-8.

Ivanov, V.V., Markusova, V.A. & Mindeli, L.E. (2016). Government investments and the publishing activity of higher educational institutions: Bibliometric analysis. *Herald of the Russian Academy of Sciences*, 86(4), 314–321. DOI: 10.1134/S1019331616040031.

Leydesdorff, L. (2000). The triple helix: an evolutionary model of innovations. *Research Policy*, 29 (2), 243–255. DOI: 10.1016/S0048-7333(99)00063-3.

National Science Board. (2012). *Diminishing Funding and Rising Expectations: Trends and Challenges for Public Research Universities, A Companion to Science and Engineering Indicators 2012*. Arlington, VA: National Science Foundation. Retrieved March 14, 2018 from: https://www.nsf.gov/nsb/publications/2012/nsb1245.pdf.

STI Conference 2018 · Leiden

Pislyakov, V. & Shukshina, E. (2014). Measuring excellence in Russia: Highly cited papers, leading institutions, patterns of national and international collaboration. *Journal of the Association for Information Science and Technology*, 65(11), 2321–2330. DOI: 10.1002/asi.23093.

Rodionov, D., Yaluner, E. & Kushneva, O. (2015). Drag race 5–100–2020 national program. *European Journal of Science and Theology*, 11(4), 199–212.

Schiermeier, Q. (2010). Russia to boost university science. *Nature*, 464(7293), 1257. DOI: 10.1038/4641257a.

Seglen, P.O. (1997). Why the impact factor of journals should not be used for evaluating research. *British Medical Journal*, 314(7079), 498–502. DOI: 10.1136/bmj.314.7079.497.

Surovitskaya, G. (2017). Comparative competitiveness of Russian flagship universities. *University Management: Practice and Analysis*, 21(4), 63–75. DOI: 10.15826/umpa.2017.04.050.

Tijssen, R.J.W., Visser, M.S. & van Leeuwen, T.N. (2002). Benchmarking international scientific excellence: Are highly cited research papers an appropriate frame of reference? *Scientometrics*, 54(3), 381–397. DOI: 10.1023/A:1016082432660.

Yuret, T. (2016). Is it easier to publish in journals that have low impact factors? *Applied Economics Letters*, 23(11), 801–803. DOI: 10.1080/13504851.2015.1109034.